\documentstyle[aps,multicol,pre,epsfig]{revtex}

\newcommand{\beq}{\begin{equation}}
\newcommand{\eeq}{\end{equation}}

\begin{document}

\title{Static and rotating domain-wall crosses in Bose-Einstein condensates}

\author{Boris A. Malomed$^{1}$, H.E. Nistazakis$^{2}$, D.J. Frantzeskakis$^{2}$, and P.G. Kevrekidis$^{3}$}

\address{
$^1$Department of Interdisciplinary Studies, Faculty of Engineering, Tel
Aviv University, Tel Aviv 69978, Israel\\
$^2$Department of Physics, University of Athens, Panepistimiopolis,
Zografos, Athens 15784, Greece\\
$^3$Department of Mathematics and Statistics, University of Massachusetts,
Amherst MA 01003-4515, USA}

\maketitle

\begin{abstract}
For a Bose-Einstein condensate (BEC) in a two-dimensional (2D)
trap, we introduce cross patterns, which are generated by
intersection of two domain walls (DWs) separating immiscible
species, with opposite signs of the wave functions in each pair of
sectors filled by the same species. The cross pattern remains
stable up to the zero value of the immiscibility parameter
$|\Delta |$, while simpler rectilinear (quasi-1D) DWs exist only
for values of $|\Delta |$ essentially exceeding those in BEC
mixtures (two spin states of the same isotope) currently available
to the experiment. Both symmetric and asymmetric cross
configurations are investigated, with equal or different numbers
$N_{1,2}$ of atoms in the two species. In rotating traps,
\textquotedblleft propellers\textquotedblright\ (stable revolving
crosses) are found too. A full stability region for of the crosses
and propellers in the system's parameter space is identified,
unstable crosses evolving into arrays of vortex-antivortex pairs.
Stable rotating rectilinear DWs are found too, at larger vlues of
$|\Delta |$. All the patterns produced by the intersection of
three or more DWs are unstable, rearranging themselves into ones
with two DWs. Optical \textquotedblleft propellers" are also
predicted in a twisted nonlinear photonic-crystal fiber carrying
two different wavelengths or circular polarizations, which can be
used for applications to switching and routing.
\end{abstract}

\vspace{2mm}

\begin{multicols}{2}

\section{Introduction}

Domain walls (DWs) separating immiscible species are generic dynamical
structures in mixed Bose-Einstein condensates (BECs) \cite{DWs}. Originally,
the DWs were studied in one-dimensional (1D) BEC models, but they can be
naturally extended into the 2D geometry as quasi-1D objects \cite{dw2,EPJD};
in particular, circular DWs, between a less repulsive component in the
middle of the trap and a more repulsive one forming an outer shell, have
been found. Recently, complex 2D large-area structures in rotating binary
BECs were predicted in simulations, including vortex lattices and sheets in
mixtures \cite{Ueda,rotTrap2comp}.

Our aim is to construct genuinely two-dimensional DW patterns, in the form
of {\it crosses} formed by intersection of two DWs, in trapped binary BECs.
We will consider both symmetric and asymmetric crosses, depending on the
ratio of the numbers of atoms in the two species (\textquotedblleft
stoichiometric ratio"). Patterns in a rotating trap will be considered too.
A rotating DW cross seems like a \textquotedblleft double-vane propeller",
which may be also symmetric or asymmetric. Note that the DWs may move, in
the general case; however, their motion is not amenable to straightforward
observation in the 1D settings, because of the relatively small size of
domains available in BEC experiments. Thus, the \textquotedblleft propeller"
offers a unique possibility to create {\em permanently moving} DWs. We
identify stability regions for the quiescent and rotating crosses, and
investigate the evolution of unstable ones. All multi-handed crosses, formed
by the intersection of more than two DWs, are shown to be unstable. We will
also briefly consider rotating one-dimensional (rectilinear) DWs, i.e.,
\textquotedblleft single-vane propellers" (however, we conclude that the
rectilinear DWs are less relevant than the crosses for experiments with
currently available BEC mixtures).

An appealing feature of the cross structures is the feasibility of their
experimental realization in binary BECs. Experimental creation of
two-component mixtures has already been reported for different spin states
in $^{87}$Rb \cite{myatt,dsh} and $^{23}$Na \cite{kett}. Accordingly, the
use of a mixture of two spin states of the same isotope seems to be the most
straightforward way of creating the DW cross configuration, as the latter
can be imprinted onto the BEC by optical beams passed through a properly
designed phase mask. Moreover, the DW cross appears, in our 
extensive computations, to be the most robust structure among the
various types of domain walls considered.
Indeed, the immiscibility condition for the repulsive BEC
mixture (the one with positive scattering lengths of atomic collisions) is
\begin{equation}
\Delta \equiv \alpha_{11}\alpha_{22}-\alpha_{12}^{2}\leq 0,  
\label{Delta}
\end{equation}
where $\alpha_{11}$, $\alpha_{22}$ and $\alpha_{12}$ are strengths of the intra- and inter-species interactions, respectively [see Eqs. (\ref{gpes}) and (\ref{gp2}) below]. As is shown below, the crosses (both
static and rotating ones) remain stable up to $|\Delta |~=0$,
while the rectilinear DW is stable only for
\begin{equation}
\Delta \leq \Delta _{{\rm rect}}^{{\rm (cr)}}=-0.061.  \label{rect-cr}
\end{equation}On the other hand, experimental measurements \cite{myatt,dsh} yield an
extremely small actual value of $|\Delta |$ in the mixture of two spin
states of $^{87}$Rb,
\begin{equation}
\Delta _{{\rm Rb}}\approx -0.0009,  \label{Rb}
\end{equation}which {\em does not} satisfy the condition (\ref{rect-cr}). For the a
mixture of different spin states in $^{23}$Na, the immiscibility parameter
extracted from experimental measurements is larger,
\begin{equation}
\Delta _{{\rm Na}}\approx -0.036  \label{Na}
\end{equation}\cite{kett}, but it does not meet the condition (\ref{rect-cr}) either.

We also note in passing that work is currently in progress towards
the creation of two-component BECs with different atomic species,
such as $^{41}$K:$^{87}$Rb \cite{KRb} and $^{7}$Li:$^{133}$Cs
\cite{LiCs}. In that case, $|\Delta |$ may take values very
different from those given in Eqs. (\ref{Rb}) and (\ref{Na}).
Furthermore, we should note that an additional possibility to push
the experimental values of $\Delta$ toward the critical one mentioned
above, is through the use of the, so called, Feshbach resonance \cite{inouye}
that can control the strength of the inter-atomic interactions.

The same \textquotedblleft propeller" effect may be implemented in
completely different physical media, namely, photonic crystals (PCs) or
photonic-crystal fibers (PCFs), with a self-defocusing nonlinearity and a
superimposed twist. In PCFs, the twist bends the inner holes, which run
along the fiber, into helices. In that case, the cross configuration may be
realized as an intersection of stable optical DWs separating two orthogonal
circular polarizations or two different carrier wavelengths (previously,
spatial-domain optical DWs between different polarizations were predicted in
driven dissipative cavities \cite{Maxi}, but no experimental setting was
proposed to observe their permanent motion). The twist-induced rotation of
the DW cross may take place along the propagation distance, so that the
entire pattern will seem like a double helix. Twisted PCs per se were
studied in a stack model \cite{stack}, and very recently, the first twisted
PCF was fabricated \cite{PCF} (it was used as a polarization filter). Note
that, combining a modulation of the air holes density along the radius of
the PCF core and a properly structured cladding, the distribution of the
effective refractive index across the PCF can be made similar to the
parabolic trapping potential in the BEC model (see below).

Besides the first possibility to observe permanent motion of
optical DWs (in the spatial domain), the proposed configuration
suggests a design of a new all-optical switch: the turn of the
sectors occupied by a signal field, with a specific circular
polarization or wavelength, just by $90^{\circ }$ is sufficient
for complete switching. The diameter of the twist-tolerant PCFs,
$\simeq 100$ $\mu $m, and their length, a few centimeters (which
is tantamount to tens of rotation periods) \cite{PCF}, make these
experiments and applications quite feasible. Moreover, the use of
the co-propagation of several different wavelengths in the PC
fiber can make {\em stable multi-handed} cross and propeller
possible. The latter may be promising for routing applications in
wavelength-division-multiplexed (WDM) telecommunications. Similar
configurations with more than two species may also be relevant in
terms of immiscible multi-component BEC mixtures. The latter issue
will be considered elsewhere.

The paper is organized as follows. In Section II we construct the
rectilinear and cross DW configurations and study some of their properties.
In Section III, we focus on the stability of crosses, including rotating
ones. The above-mentioned \textquotedblleft single-vane
propeller" (a rotating rectilinear DW)\ are also discussed.
Section IV summarizes our findings and presents our conclusions.

\section{Construction and properties of domain-wall configurations}

The two-component rotating repulsive BEC is described by a system of coupled
Gross-Pitaevskii (GP) equations \cite{fetter},
\begin{equation}
i\hbar \frac{\partial \psi _{j}}{\partial t}=\left[
\hat{H}+\sum_{k=1,2}g_{jk}|\psi _{k}|^{2}\right] \psi _{j},\,\,\,\
j=1,2, \label{gpes}
\end{equation}where $\psi _{j}$ are the wave functions of the two species, normalized so
that $N_{j}=\int |\psi _{j}|^{2}d{\bf r}$ is the respective number
of atoms. The single-species Hamiltonian in Eq. (\ref{gpes}) is
$\hat{H}=-(\hbar ^{2}/2m)\nabla ^{2}-\omega
_{L}\widetilde{L}_{z}+\widetilde{V}$, where $m$ is the atomic mass
(assuming a mixture of two spin states of the same isotope),
$\omega _{L}$ is the rotation frequency, and
$\widetilde{L}_{z}=i\hbar (x\partial _{y}-y\partial _{x})$ is the
angular-momentum operator. The trapping potential is
\begin{equation}
\widetilde{V}=\frac{m}{2} (\omega _{r}^{2}r^{2}+\omega _{z}^{2}z^{2}),
\label{trap}
\end{equation}
where $r^{2}\equiv x^{2}+y^{2}$, and the confining frequencies
$\omega _{r}$ and $\omega _{z}$ are assumed to obey the condition
$\omega _{r}/\omega _{z}\equiv \Omega \ll 1$. The intra- and
inter-species interactions are characterized by the coefficients
$g_{jj}=4\pi \hbar ^{2}a_{jj}/m$ and $g_{12}\equiv g_{21}=4\pi
\hbar ^{2}a_{12}/m$, respectively, where $a_{jk}$ are the
corresponding scattering lengths; as mentioned above, we consider the
(most typical) case of positive $a_{jk}$. Then, the condition of
the immiscibility between the components takes the well-known form
of Eq. (\ref{Delta}).

Following Refs. \cite{GPE1d}, effective 2D GP equations can be
derived from the 3D ones. To this aim, measuring the coordinates
and time in units of the harmonic-oscillator length and period,
i.e., $\sqrt{\hbar /m\omega _{z}}$ and $1/\omega _{z}$,
respectively, we seek for solutions to Eqs. (\ref{gpes}) as $\psi
_{j}(r,z,t)=(2\pi )^{-1/4}\sqrt{\hbar \omega
_{z}/g_{11}}u_{j}(r,t)\Phi _{j}(z)\exp \left( -i\gamma
_{j}t\right) $, where $\Phi _{j}(z)=\pi ^{-1/4}\exp (-z^{2}/2)$ is
the ground state of the 1D harmonic oscillator. Multiplying the
resulting equations by $\Phi ^{\star }$ and integrating it in $z$,
we arrive at a system of 2D equations,
\begin{eqnarray}
i\frac{\partial u_{j}}{\partial t} &=&\left[ \hat{H}_{{\rm 2D}}+\sum_{k=1,2}\alpha_{jk}|u_{k}|^{2}\right] u_{j},\,\,\,\ j=1,2,  \label{gp2} \\
\hat{H}_{{\rm 2D}} &\equiv &(-1/2)\nabla _{\perp }^{2}+V(r)-\Omega _{L}L_{z},
\label{H}
\end{eqnarray}
where $\nabla _{\perp }^{2}$ is the 2D Laplacian, and
\begin{equation}
V(r)\equiv (1/2)\Omega ^{2}r^{2},L_{z}\equiv i(x\partial _{y}-y\partial
_{x}),\Omega _{L}\equiv \omega _{L}/\omega _{z}.  
\label{V}
\end{equation}
The nonlinearity coefficients in the $2$D system (\ref{gp2}) are $\alpha_{11}=1$, $\alpha_{12}=g_{12}/g_{11}$, and $\alpha_{22}=g_{22}/g_{11}$.
Then, the
numbers of atoms in the species are $N_{j}=\left( 4\sqrt{2}\pi
^{3/2}\right) ^{-1}\sqrt{\hbar /m\omega _{z}}Q_{j}$, where
$Q_{j}\equiv \int |u_{j}|^{2}d^{2}{\bf r}_{\perp }$ are the norms
of the 2D wave functions.

The spatial evolution of bimodal optical beams in the above-mentioned
twisted PCFs is also described by Eqs. (\ref{gp2}) for the amplitudes $u_{j}$
of the two components, with $t$ replaced by the propagation distance $z$,
and $V(r)$ in Eq. (\ref{H}) is a potential function describing the cladding
surrounding the PCF proper. In this case, $N_{j}$ are total powers of the
two components of the beam, and the nonlinear coefficients take values close
to $\alpha_{12}=2\alpha_{11}=2\alpha_{22}$; according to Eq. (\ref{Delta}), the latter
values definitely guarantee the \textquotedblleft immiscibility" (mutual
repulsion of the two optical modes).

Solutions to Eqs. (\ref{gp2}) in the form of either a simple rectilinear DW
(first, without the rotation, $\Omega_{L}=0$) are constructed as follows.
We start with the Thomas-Fermi (TF) configuration for the two components
with chemical potentials $\mu _{j}$,
\begin{equation}
\left( u_{j}\right) _{{\rm TF}}=e^{-i\mu _{j}t}\sqrt{\left[ \mu
_{j}-V(r)\right] /\alpha_{jj}},  \label{TF}
\end{equation}
if $\mu _{j}-V(r)>0$, and $\left( u_{j}\right) _{{\rm TF}}=0$ otherwise (in
the PCFs, $-\mu _{j}$ are the propagation constants of the optical modes).
Then, to construct a rectilinear DW, we effectively remove each component
from a half-plane, upon multiplying $\left( u_{j}\right) _{{\rm TF}}$ by
\begin{equation}
f_{j}(x)\equiv \frac{1}{2} \left[1-(-1)^{j}\tanh x \right].  \label{f}
\end{equation}
The resulting function $u_{1}$ ($u_{2}$) is 
positive in the
left (right) half of the $\left( x,y\right) $ plane.

To construct a DW cross in a similar fashion, we also start from the TF
ansatz (\ref{TF}), and then remove two quarter-planes in each component,
multiplying the expression (\ref{TF}) by $\left( 1/2\right) \left[
f_{j}(x)f_{j}(y)+f_{j}(-x)f_{j}(-y)\right] $, where the factors $f_{1,2}$
are defined as per Eq. (\ref{f}). As a result, the wave functions $u_{1}$
and $u_{2}$ have support, respectively, practically 
only in the quadrants (sectors) $1$
and $3$ (i.e., $xy>0$), and $2$ and $4$ (i.e., $xy<0$). Next, we multiply
the entire configuration by $(-1)^{j}\tanh x$, hence the functions $u_{1}$
and $u_{2}$ are, respectively, positive (negative) in the sectors $1$ ($3$)
and $2$ ($4$). In other words, we introduce the phase shift of $\pi $
between the two quadrants filled by each species (in other words, we aim to
construct a DW cross which, simultaneously, is a \textquotedblleft latent
dark soliton" in each component). Without these phase shifts, the cross will
be obviously unstable against splitting into a set of two approximately
parallel quasi-rectilinear DWs, as a pair of the sectors filled by the same
species without the phase shift between them will tend to merge into a
single stripe-like pattern.

After the prototype rectilinear or cross DW configuration was constructed as
described above, Eqs. (\ref{gp2}) were numerically integrated in imaginary
time, to let the system relax into a stable stationary state closest to the
prototype one (which is a well-known technical ruse). For the DW-cross case,
the relaxation in imaginary time did not essentially alter the prototype
pattern, always converging to a numerically exact stable cross pattern,
provided that the immiscibility condition (\ref{Delta}) was met. This
outcome of the imaginary-time integration took place for {\em arbitrarily
small} values of the immiscibility coefficient $|\Delta |$, including the
case when $\Delta $ was set precisely equal to 0. With $\Delta >0$, the
integration always showed that two species would completely mix into a
uniform state.

In the case of the rectilinear DW, the imaginary-time integration
converged to a stable quasi-1D DW pattern only if the condition
(\ref{rect-cr}) was satisfied, which is more restrictive than the
general immiscibility criterion (\ref{Delta}). For smaller values
of $|\Delta |$, stable rectilinear DWs could never be found;
instead, the two species would completely mix up, even if $\Delta
$ was negative. 
Thus, the DW-cross
pattern, although being more complex than its simplest rectilinear
counterpart, is a much more robust one, and has a much better
chance to be created in experiments with the currently available
{\em weakly immiscible} spin-state mixtures that actually do not
satisfy the condition (\ref{rect-cr}), see Eqs. (\ref{Rb}) and
(\ref{Na}). The extra robustness of the DW-cross structure may be
explained by the additional $\pi $ phase shifts lent to its both
components, as described above.

As an example, in Fig.1 we show the result of the relaxation for each type
of the prototype pattern, with the value of $\Delta $ taken as for the
spin-state mixture in $^{87}$Rb, see Eq. (\ref{Rb}). This was realized by
setting $\alpha_{11}=1$, $\alpha_{22}=0.94$, and $\alpha_{12}=0.97$ in Eq. (\ref{gp2}). As
is seen, the rectilinear DW pattern cannot be formed indeed, preferring to
mix itself up into a uniform state [panels (a) and (b)]. On the contrary to
that, the initial (prototype) DW-cross pattern readily relaxes into a
configuration of exactly the same type.

In the example shown in Fig. 1, the normalized magnetic-trap
strength is $\Omega =0.05$ [see Eq. (\ref{V})], and both chemical
potentials are set equal to $1$. In terms of the real-world
parameters, this choice corresponds to a mixture of two spin
states in the $^{87}$Rb condensate in a disk-shaped trap with
\begin{equation}
\omega _{r}=2\pi \times 6~{\rm Hz},~\omega _{z}=2\pi \times 120~{\rm Hz}
\label{omegaomega}
\end{equation}
[see Eq. (\ref{trap})], the initial TF radius and numbers of atoms in each
species being
\begin{equation}
R=34\mu {\rm m},~N=3.6\times 10^{3}.  \label{RN}
\end{equation}

In what follows below, we will display examples for larger $|\Delta |$,
which may be unrealistically large directly (but may be attainable
through the use of the Feshbach resonance \cite{inouye} mentioned 
above). This allows
us to generate patterns that are much sharper and easier to understand,
while qualitatively they are completely tantamount to those found at smaller
(more directly realizable) values of the immiscibility parameter. So, we will
fix
\begin{equation}
\alpha_{11}=1, \,\,\,\, \alpha_{22}=1.01, \,\,\,\, \alpha_{12}=1.52,  
\label{aaa}
\end{equation}which corresponds to $\Delta =-1.3$.

\begin{figure}[tbp]
\centering \includegraphics[width=3.5in]{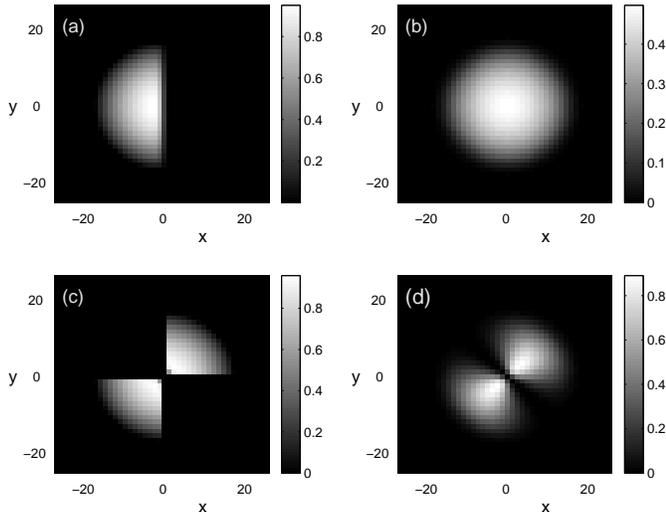}
\caption{Gray-scale plots showing the distribution of the density
$|u_{1}|^{2}$ of one species (the density $|u_{2}|^{2}$ is
complementary to $|u_{1}|^{2}$) in the initial configuration (a)
corresponding to the prototype rectilinear DW, and in the
corresponding final counterpart (b), which is generated by the
numerical integration of the GP equations (\protect \ref{gp2}) in
imaginary time, in the case of weak immiscibility, with $\Delta
=-9 \times 10^{-4}$ (corresponding to $\alpha_{11}=1$, $\alpha_{22}=0.94$, $\alpha_{12}=0.97$) and $\Omega=0.05$. The panels (c) and (d) display the same for the initial
and final configurations in the case of the DW cross.}
\label{Fig1}
\end{figure}

\begin{figure}[tbp]
\centering \includegraphics[width=3.5in]{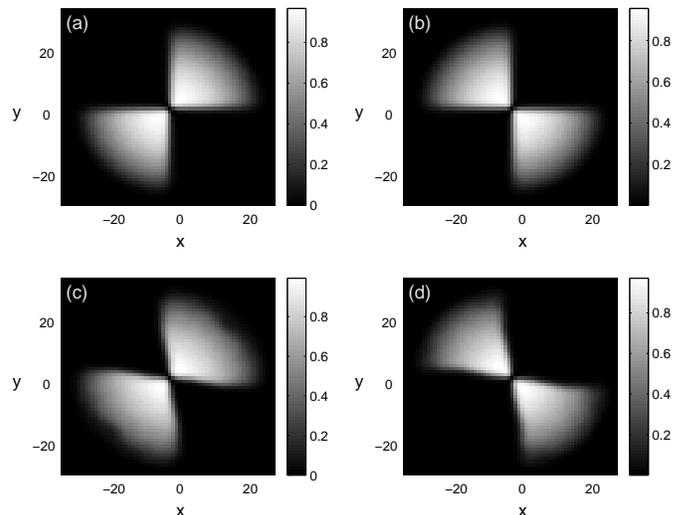}
\caption{Gray-scale plots of the densities of the two species,
$|u_{1}|^{2}$ and $|u_{2}|^{2}$ (left and right panels), in the
symmetric (a, b) and asymmetric (c, d) DW-cross patterns. The
parameters are $\alpha_{11}=1$, $\alpha_{22}=1.01$, $\alpha_{12}=1.52$, and
$\Omega =0.05$. In the symmetric configuration, $\protect\mu
_{1}=\protect\mu _{2}=1$, and in the asymmetric one $\protect\mu
_{1}=1.1025$, $\protect\mu _{2}=1$, and $N_{1}/N_{2}=1.2$,
$\protect\theta _{1}=6\protect\pi /11$, $\protect\theta
_{2}=5\protect\pi /11$. The estimate (in physical units) for the Thomas-Fermi
diameter for both configurations is $68 \protect\mu $m.}
\label{Fig2}
\end{figure}

Before proceeding further, it is important to note that, alongside
the symmetric DW-cross configurations like the one shown in Fig.
1(d), {\it asymmetric} ones are possible too, with different
numbers of atoms in the two species, $Q_{1}\neq Q_{2}$. Indeed,
the TF approximation (\ref{TF}) yields
\begin{equation}
Q_{j}=\left( \frac{\pi }{2\alpha_{jj}}\right) \left( \mu _{j}-\frac{1}{4}\Omega
^{2}R^{2}\right) R^{2},  
\label{Qj}
\end{equation}
where $R$ is the TF radius of the state; consequently, $\alpha_{11}\neq \alpha_{22}$
and/or $\mu _{1}$ $\neq \mu _{2}$ lead to $Q_{1}\neq Q_{2}$, and, as a 
result, to asymmetric crosses. We have checked that the numerical results
always obey a natural relation, $N_{1}/N_{2}\equiv Q_{1}/Q_{2}=\theta
_{1}/\theta _{2}$, where $\theta _{j}$ is the intrinsic angle of the $j$-th
sector, with $\theta _{1}+\theta _{2}\equiv \pi $. Examples of the symmetric
and asymmetric DW crosses are shown, respectively, in the top and bottom
panels of Fig. \ref{Fig2}.

\section{Stability and rotation of the domain-wall patterns}
\subsection{Static domain-wall crosses}

\begin{figure}[tbp]
\centering \includegraphics[width=1.6in]{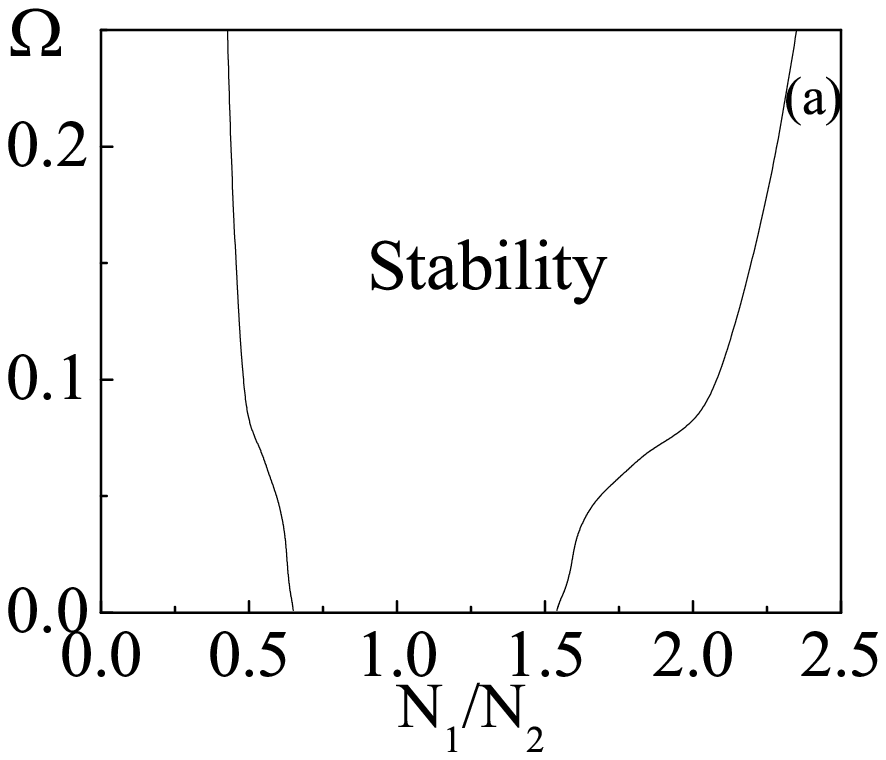} \centering
\includegraphics[width=1.6in]{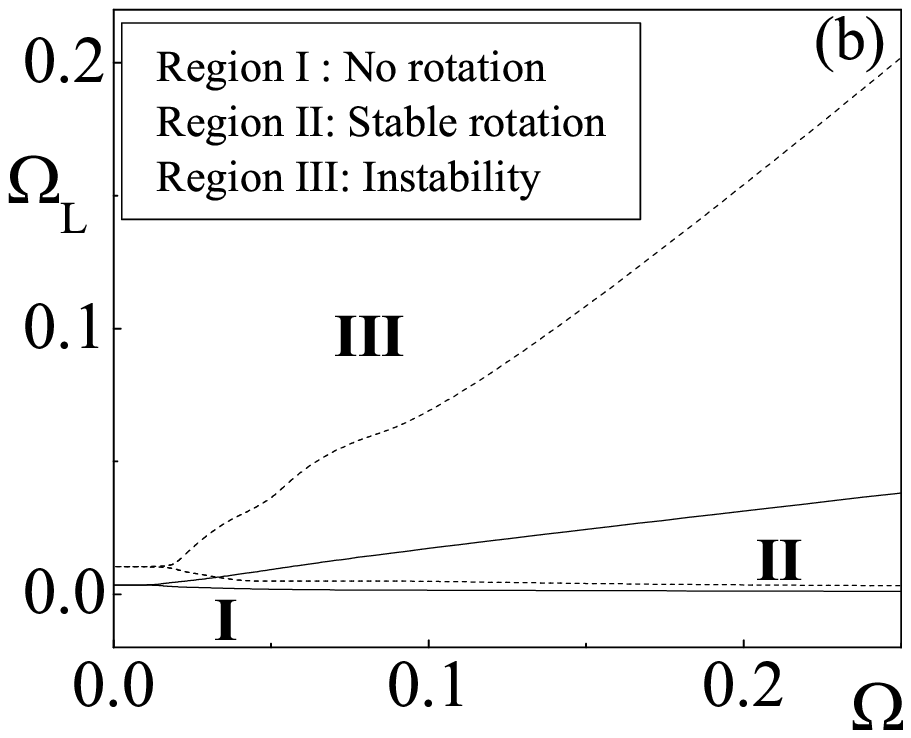}
\caption{Stability regions for the static (a) and rotating (b) DW
crosses. The values of the nonlinear coefficients are fixed as in
Eqs. (\protect\ref{aaa}). In (b), solid and dashed lines,
respectively, show borders of the stable-rotation region for
symmetric and asymmetric crosses (in the latter case,
$N_{1}/N_{2}=1.5$).} \label{Fig3}
\end{figure}

Dynamical stability of the symmetric and asymmetric DW crosses is
a crucially important issue. We have found that, for a fixed
trap's strength $\Omega $, the stability strongly depends on the
stoichiometry ratio $N_{1}/N_{2}$, more symmetric configurations
being more robust. To show this, we fix the nonlinearity
coefficients $\alpha_{jj}$ as in Eqs. (\ref{aaa}), 
we set $\mu_{2}=1$ and, then, for a given value of $\Omega$ in the interval $(0,0.25)$, we vary $\mu_{1}$ to induce variation of the ratio $N_{1}/N_{2}\equiv Q_{1}/Q_{2}$, as per Eq. (\ref{Qj}). Finally, we simulate Eqs. (\ref{gp2}) in real time (up to $t=1000$), to test
the stability of the configuration. The resulting stability domain in the $(N_{1}/N_{2},\Omega )$ parametric plane is displayed in Fig. \ref{Fig3}(a). For example, at $\Omega =0.05$, which
corresponds to the above-mentioned values (\ref{omegaomega}) and (\ref{RN}) of the physical parameters, Fig. \ref{Fig3}(a) shows that stable DW crosses exist in the interval $0.65\leq
N_{1}/N_{2}\leq 1.65$. We note that, alternatively, it is possible to represent the domain of stability of the static DW crosses in the $(\mu_{1}, \mu_{2})$ plane, for a fixed value of $\Omega$ (then, each point of the domain would correspond to a given value of $N_{1}/N_{2}$). This can be done, e.g., upon rescaling the variables in Eqs. (\ref{gp2})-(\ref{H}) so that $\Omega=1$ (see also a relevant discussion in Section III.B); in that case the domain of stability would be a triangular-like zone enclosing the line $\mu_{1}=\mu_{2}$. 

\begin{figure}[tbh]
\centering
\includegraphics[width=3in]{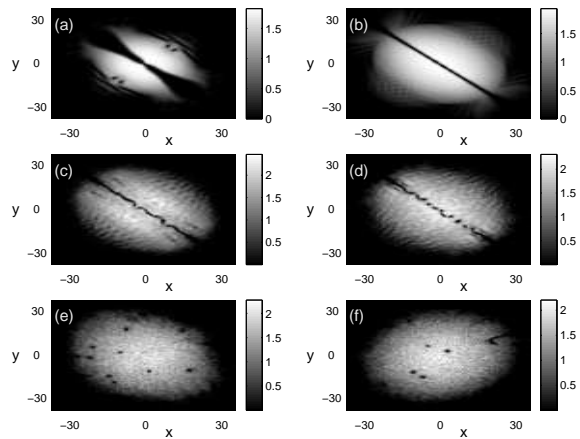}
\caption{Evolution of the density $|u_{1}|^{2}$ of the species
with the larger number of atoms in an unstable DW cross, for the
same parameters as in Fig. \protect\ref{Fig2}, but with
$\protect\mu _{1}=1.8025$. In this case, $N_{1}/N_{2}=2.5$ and
$\protect\theta _{1}=5\protect\pi /7$, $\protect\theta
_{2}=2\protect\pi /7$. The snapshots (a)-(f) correspond to $t=0$,
$t=100$, $t=310$, $t=316$, $t=1000$ and $t=2200$, respectively
(the physical time unit is $1.33$ ms). The density $|u_{2}|^{2}$
is complementary to $|u_{1}|^{2}$, as in Figs.
\protect\ref{Fig2}(c, d).} 
\label{Fig4}
\end{figure}

Evolution of unstable DW crosses is also an issue of interest. A
typical example of the instability development (in the component
with the larger number of atoms) is displayed in Fig. \ref{Fig4}
for $\mu _{1}=1.8025$ and $\mu _{2}=1$, which corresponds to
$N_{1}/N_{2}=2.5$. At an initial stage, the cross quickly
rearranges into a quasi-1D object, which is actually a dark
soliton in the species with the larger number of atoms, coupled to
a bright soliton in the other species (this object resembles
structures considered in Ref. \cite{EPJD}). The latter
configuration is itself subject to a {\it snaking instability},
which is a known feature of quasi-1D dark solitons in BECs filling
2D areas \cite{thsnbec} (we stress that, unlike the dark solitons,
the rectilinear DWs are not subject to this instability). The
snaking instability initiates break-up of the dark stripe into
four vortex-antivortex pairs in the first species, with the second
species collecting itself into spots coupled to the pairs.
Finally, the four vortex-antivortex pairs annihilate into two. The
latter configuration persists for long times, and it may be
related to serpentine-shaped vortex sheets reported in recent 2D
simulations of large-size BECs \cite{Ueda}. It would be desirable
to analyze this instability by means of a finite mode stability analysis
in the form of \cite{perez}. However, in the present setting, the 
modulus of the wavefunction is not radially symmetric (nor does
the DW configuration bear a complex phase structure), hence it is not 
directly amenable to such an approach. Understanding the origin of
such an instability would consitute an interesting topic for future
studies.

\subsection{Domain-wall crosses in the rotating trap}

For the rotating trap, with $\Omega _{L}\neq 0$ in Eqs. (\ref{gp2}), we have
found that both the symmetric and asymmetric DW crosses revolve stably in a
certain frequency interval,
\begin{equation}
\left( \Omega _{L}\right) _{\min }<\Omega _{L}<\left( \Omega _{L}\right)
_{\max }.  \label{interval}
\end{equation}
If $\Omega _{L}<\left( \Omega _{L}\right) _{\min }$, the cross does not
rotate at all, while at $\Omega _{L}>\left( \Omega _{L}\right) _{\max }$ it
decays (see below). These results are summarized in Fig. \ref{Fig3}(b). In
region I, the DW cross remains quiescent, in region III it gets destroyed,
while the stable rotation occurs in region II. Note that there are minimum
values of the trap strength which are necessary for the rotation: $\Omega
_{\min }=0.012$ and $0.018$ for the symmetric and asymmetric
\textquotedblleft double-vane propellers". Those values correspond to the
edge points on the solid and dashed lines, respectively, in Fig. 3(b), the
rotation frequencies at these points being $\Omega _{L}=0.0034$ and $0.010$.

Note that the horizontal axis in Fig. 3(b) is extended up to a relatively
large value, $\Omega =0.25$ (recall that it was assumed above that the
magnetic-trap's strength $\Omega $ is a small parameter, in order to derive
the 2D GP equations from the underlying 3D system). According to experience
accumulated in applications of asymptotic methods to various models
(including the GP equation), the value $0.25$ is small enough to produce
reliable results.

\begin{figure}[tbp]
\centering
\includegraphics[width=3in]{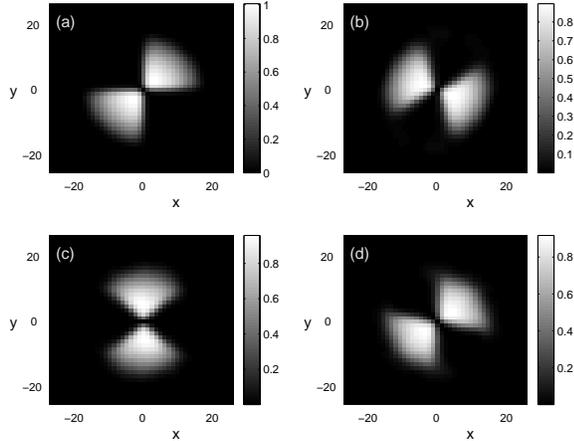}
\caption{A stable rotating assymetric domain-wall cross (\textquotedblleft
double-vane propeller"). The four snapshots (a), (b), (c), and (d)
show the distribution of the density $|u_{1}|^{2}$ at consecutive
time moments -- respectively, $t=0$, $T/6$, $T/3$, and $T/2$,
which cover a half of the rotation period, $T/2=\protect\pi
/\Omega _{L}\approx 52.3$ ($\simeq 70$ ms, in physical units). The
parameters are as in Fig.\ref{Fig2}(c, d) but with $\Omega=0.1$ and  
$\Omega_{L}=0.06$.} 
\label{Fig5}
\end{figure}

\begin{figure}[tbh]
\centering \includegraphics[width=3in]{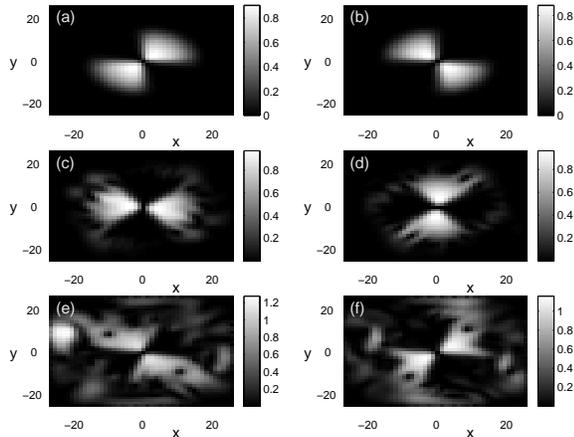}
\caption{An example of the evolution of an unstable symmetric 
\textquotedblleft double-vane propeller" (rotating domain-wall
cross). The panels in the left and right columns display the
density distributions $|u_{1}|^{2}$ and $|u_{2}|^{2}$,
respectively, at $t=0$ [frames $(a)$ and $(b)$], $t=10$ [frames
$(c)$ and $(d)$], and $t=20$ [frames $(e)$ and $(f)$]. The
parameters are as in Fig. \protect\ref{Fig2}(a,b), but with
$\Omega =0.1$ and $\Omega _{L}=0.08$.} 
\label{Fig6}
\end{figure}

Typical values of the physical parameters admitting the stable
rotation of the DW cross can again be estimated for a mixture of
two spin states in the $^{87}$Rb condensate. For instance, taking
the trapping frequencies $\omega _{r}=2\pi \times 7$ Hz and
$\omega _{z}=2\pi \times 70$ Hz, the TF radius $R=22~\mu $m, and
$10^{3}$ atoms in each species, 
the rotation frequency $\omega _{L}=2\pi \times 
0.84$ Hz definitely falls within the stable-rotation interval
(\ref{interval}). Examples of a stably rotating propeller with
$N_{1}/N_{2}=1.2$, and of its self-destruction in the case of
$\Omega _{L}>\left( \Omega _{L}\right) _{{\rm max}}$ (for
$N_{1}=N_{2}$) are shown in Figs. \ref{Fig5} and \ref{Fig6},
respectively.

\begin{figure}[tbp]
\centering\includegraphics[width=3in]{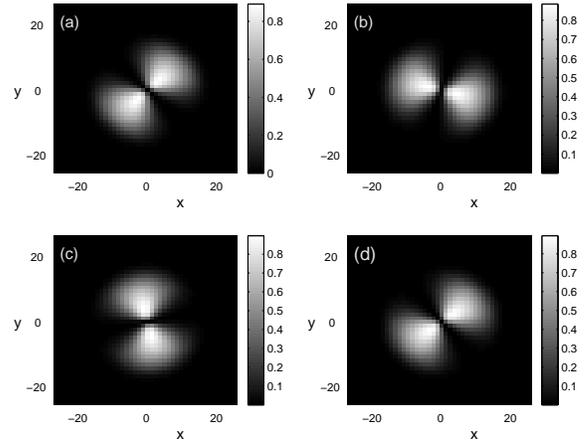}
\caption{A stable rotating symmetric domain-wall cross found exactly at the
immiscibility border, $\Delta =0$ (corresponding to $\alpha_{11}=\alpha_{22}=\alpha_{12}=1$) and $\Omega=0.1$. The arrangement of the figure is the same 
as in Fig. \protect\ref{Fig5}.}
\label{Fig:prop_rot_dwc}
\end{figure}

As it was mentioned above, a dynamical property of the DW crosses
which is crucially important for their relevance to the currently
available experimental settings, that have a very small value of
the immiscibility parameter $|\Delta |$ [see Eq. (\ref{Delta})],
is the fact that the crosses remain stable for all the negative
values of $\Delta $, up to $\Delta =0$. This robustness carries
over to the rotating crosses, as illustrated by Fig.
\ref{Fig:prop_rot_dwc}, that shows a stable rotating cross in the
case of $\Delta =0$.

\begin{figure}[tbh]
\centering \includegraphics[width=3.5in]
{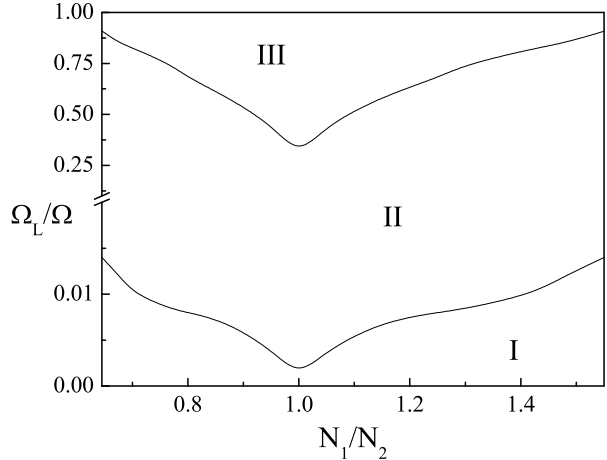}
\caption{The region of the stable rotation of the domain-wall crosses, shown in the parameter plane $\left(N_{1}/N_{2},\Omega_{L}/\Omega \right)$. In region I, the DW cross is quiescent, in region III it is destroyed, while in region II it is rotating in a stable manner. Note that for values of $N_{1}/N_{2}$ outside the interval $(0.65, 1.55)$ the DW crosses are unstable.
}
\label{Fig7}
\end{figure}

The results can alternatively be summarized upon considering a rescaling of variables in Eqs. (\ref{gp2})-(\ref{H}). In particular, upon measuring the time, spatial variables and normalized wavefunctions in units of $\Omega^{-1}$, $\Omega^{-1/2}$ and $\Omega^{1/2}$ respectively, we obtain a system of two equations similar to Eq. (\ref{gp2}), but with the Hamiltonian being given by,
\begin{eqnarray}
\hat{H}_{{\rm 2D}} \equiv \frac{1}{2} \nabla_{\perp }^{2}+\frac{1}{2}r^{2}-\frac{\Omega_{L}}{\Omega}L_{z}.
\label{HN}
\end{eqnarray}
Then, we may present the domain of stability of the rotating DW crosses in the parameter plane $\left(N_{1}/N_{2},\Omega_{L}/\Omega \right)$ as follows: We set $\mu_{2}=1$ and, then, for a given value of $\mu_{1}$ (which sets the value of $N_{1}/N_{2}$), we vary the ratio $\Omega_{L}/\Omega$ in the interval $(0, 1)$. Then, we numerically integrate the system of equations for $u_{j}$, for long times, to test the stability of the rotating DW cross. The results, are shown in Fig. \ref{Fig7}, where the stable-rotation region is delineated in the {\em full} parameter plane: In region I, the DW cross remains quiescent, in region III it is destroyed, while the stable rotation occurs in region II. Note that for values of $N_{1}/N_{2}$ outside the shown interval, $(0.65, 1.55)$, the DW crosses are unstable. Thus, for instance, at a typical value of the ratio between the rotation frequency and the trap strength, $\Omega_{L}/\Omega=0.1$, the stoichiometry ratio must belong to the interval $0.65\leq N_{1}/N_{2}\leq 1.55$. As mentioned above, the smallest value of $\Omega_{L}/\Omega$ necessary for the stable rotation (which is $0.0019$) corresponds to the symmetric case, $N_{1}/N_{2}=1$. 

\subsection{More general cross and propeller patterns}

\begin{figure}[tbh]
\centering \includegraphics[width=3in]{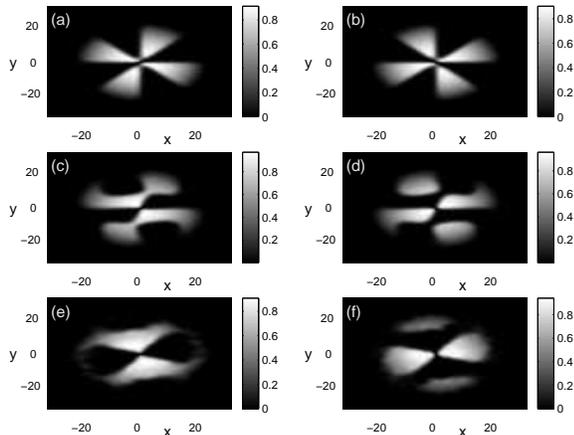}
\caption{Instability of a higher-order domain-wall cross (left and 
right panels show distributions of the densities $|u_{1}|^{2}$ and 
$|u_{2}|^{2}$). The frames (a) and (b) display the initial 
configuration at $t=0$. Further frames show the evolving
configurations: (c), (d) at $t=50$, and at (e), (f) at $t=600$.
The parameters are as in Fig. \protect\ref{Fig2}(a, b).} 
\label{Fig8}
\end{figure}

All the patterns produced by the intersection of more than two DWs
were found to be unstable, rearranging themselves into the
fundamental crosses considered above. A typical example of that,
with four intersecting DWs and $N_{1}/N_{2}=1$, is displayed in
Fig. \ref{Fig8}. The same instability of the higher-order crosses
takes place in the rotating trap.

Two-dimensional patterns similar to the symmetric DW crosses can
be formed in{\em \ single-component} BECs by two linear dark
solitons intersecting under the right angle (provided that the
dark soliton itself may be stable). We have checked that such
single-component patterns are {\em always} unstable.

The crosses in BECs (but not in the above-mentioned photonic-crystal media)
can be extended into the 3D case, as patterns formed by the intersection of
planar DWs. Moreover, torque can be applied to such a structure, making it
look like a double helix. Further, these patterns may revolve together with
the trap. Results for the cross-shaped 3D patterns will be reported
elsewhere.

\subsection{Rotating rectilinear domain walls}

\begin{figure}[tbh]
\centering\includegraphics[width=3in]{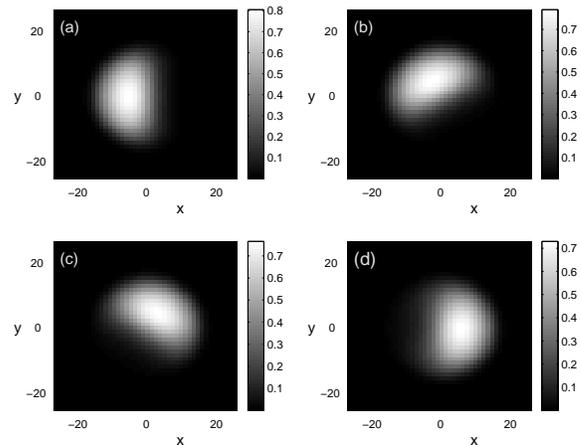}
\caption{Rotation of the rectilinear domain wall at the smallest
value of the immiscibility parameter, $|\Delta |=0.061$ (corresponding to $\alpha_{11}=\alpha_{22}=1$, $\alpha_{12}=1.03$), and $\Omega=0.1$ at which it is stable. The arrangement of the figure is the same as in Fig.\protect\ref{Fig:prop_rot_dwc}.} \label{Fig:prop_rot_rdw}
\end{figure}

As  was explained above, stable rectilinear DWs are less relevant the
currently (experimentally) tractable
small values of the immiscibility parameter $|\Delta|$. 
Nevertheless, for completeness of the analysis, and also for the
sake of possible future experiments in BEC mixtures of different atomic
species, where $|\Delta |$ may be much larger than in the presently
available mixtures of different spin states of the same isotope, it makes
sense to briefly consider a possibility of stable rotation of the DWs of
this type too. Numerical simulations show that, generally, the rectilinear
DW withstands rotation, within a certain interval of the angular velocities
[cf. Eq. (\ref{interval})], if it is stable in the static trap. Here, we
do not aim to produce comprehensive results for the stability domain of the
rotating rectilinear DWs. Instead, in Fig. \ref{Fig:prop_rot_rdw} we display
an example of a stably rotating DW of this type. This example has a special
purport, as it shows the rotating rectilinear DW found at the {\em minimum
value} of the immiscibility parameters at which it may be stable, $|\Delta
|=0.061$. Stable rotating rectilinear DWs found in the generic case are
quite similar to the one shown in Fig. \ref{Fig:prop_rot_rdw}.

\section{Conclusions}

Our analysis predicts the existence of symmetric and asymmetric domain-wall
(DW) crosses in immiscible mixtures of repulsive BECs in disk-shaped traps.
The stable crosses are distinguished by the phase shift of $\pi $ between
sectors filled by each component. The same patterns may also be carried by
bimodal light beams in photonic crystal fibers (PCFs). The crosses are
stable in a wide range of parameters, including a certain interval of the
stoichiometry ratio $N_{1}/N_{2}$, the case $N_{1}=N_{2}$ being the most
robust one. Adding the trap's rotation (or twist of the PCF, in the optical
model), we have shown that the DW crosses can rotate, provided that the
driving angular velocity (or the twist pitch, in the PCF) is limited from
below and from above (if it is too small, the cross does not rotate, and if
it is too large, the cross decays). In the PCF, the rotation takes place not
in time, but in the spatial domain, i.e., along the propagation distance,
hence the optical cross actually looks like a double helix. The rotating
optical crosses have potential for the use in switching and routing
applications. Full stability domains for the quiescent and rotating crosses
were identified, corresponding to values of physical parameters accessible
in current experiments (both for BECs and PCFs).

The crosses (including the rotating ones) persist exactly up to the zero
value of the immiscibility parameter $|\Delta |$. This aspect of the
robustness of the DW crosses is crucially important because, in the BEC
mixtures of different spin states in $^{87}$Rb and $^{23}$Na, currently
available to the experiment, the actual value of $|\Delta |$ is very small.
We have demonstrated that, in the cases of practical interest, the simplest
rectilinear (quasi-1D) DWs {\em do not exist} (in a stable form), while the
crosses remain entirely stable. For the completeness' sake, we have also
demonstrated a possibility of stable rotation of the rectilinear DWs at
larger values of $|\Delta |$.

The patterns produced by the intersection of more than two DWs were observed 
to be unstable. They rearrange themselves into the fundamental crosses.

Finally, it is worthy to mention that, at angular velocities of the rotating
trap essentially exceeding the range of values dealt with in this paper, one
may expect the existence of rotating DWs separating not merely areas with
the quasi-uniform distribution of the densities of immiscible species, but
rather areas filled with triangular vortex lattices (each species supporting
its own lattice). In single-component BECs, such regular lattices, composed
of a large number of vortices, have been recently studied in detail in
direct experiments \cite{JILA}. In fact, creation of the DW separating the
vortex lattices in immiscible components is the most straightforward way for
experimental observation a {\em rotating} DW pattern in BECs. This
possibility will be considered in detail elsewhere.

We appreciate valuable discussions with L. Carr, E. Cornell and P. Engels.
B.A.M. acknowledges a partial support from the Israel Science Foundation
through the grant No. 8006/03, and hospitality of JILA (Boulder, Colorado).
P.G.K. acknowledges support from the Eppley Foundation, NSF CAREER and
NSF-DMS-0204585.


\end{multicols}


\begin{references}
\bibitem{DWs} M. Trippenbach, K. Goral, K. Rzazewski, B.A. Malomed, and
Y.B. Band, J. Phys. B {\bf 33}, 4017 (2000);
P. \"{O}hberg and L. Santos, Phys. Rev. Lett. {\bf 86} 2918 (2001);
S. Coen and M. Haelterman, Phys. Rev. Lett. {\bf 87}, 140401 (2001);
D.T. Son and M.A. Stephanov, Phys. Rev. A {\bf 65}, 063621 (2002);
J.J. Garcia-Ripoll, V.M. P\'{e}rez-Garc\'{\i}a, and F. Sols,
Phys. Rev. A {\bf 66}, 021602 (2002);
B. Deconinck, J.N. Kutz, M.S. Patterson, and B.W. Warner,
J. Phys. A: Math. Gen. {\bf 36} 5431 (2003);
P.G. Kevrekidis, B.A. Malomed, D.J. Frantzeskakis, and A.R. Bishop,
Phys. Rev. E {\bf 67}, 036614 (2003).

\bibitem{dw2} T.L. Ho and V.B. Shenoy, Phys. Rev. Lett {\bf 77}, 3276 (1996);
B.D. Esry, C.H. Greene, J.P. Burke, Jr., and J.L. Bohn, Phys. Rev. Lett.
{\bf 78}, 3594 (1997);
P. \"{O}hberg and S. Stenholm, Phys. Rev. A {\bf 57}, 1272 (1998);
B.D. Esry and C.H. Greene, Phys. Rev. A {\bf 59}, 1457 (1999);
A.A. Svidzinsky and S.T. Chui, Phys. Rev. A {\bf 68}, 013612 (2003).

\bibitem{EPJD} P.G. Kevrekidis, H.E. Nistazakis, D.J. Frantzeskakis,
B.A. Malomed, and R. Carretero-Gonzalez,
Eur. Phys. J. D {\bf 28}, 181, (2004).

\bibitem{Ueda} K. Kasamatsu, M. Tsubota, and M. Ueda, Phys. Rev. Lett.
{\bf 91}, 150406 (2003).

\bibitem{rotTrap2comp} S. Ghosh, M.V.N. Murthy, S. Sinha, Phys. Rev. A
{\bf 64}, 053603 (2001).

\bibitem{myatt} C.J. Myatt, E.A. Burt, R.W. Ghrist, E.A. Cornell,
and C.E. Wieman, Phys. Rev. Lett. {\bf 78}, 586 (1997).

\bibitem{dsh} D.S. Hall, M.R. Matthews, J.R. Ensher, C.E. Wieman,
and E.A. Cornell, Phys. Rev. Lett. {\bf 81}, 1539 (1998)

\bibitem{kett} D.M. Stamper-Kurn, M.R. Andrews, A.P. Chikkatur,
S. Inouye, H.-J. Miesner, J. Stenger, and W. Ketterle,
Phys. Rev. Lett. {\bf 80}, 2027 (1998).

\bibitem{KRb} G. Modugno, G. Ferrari, G. Roati, R.J. Brecha, A. Simoni,
and M. Inguscio, \newblock Science {\bf 294},
1320 (2001).

\bibitem{LiCs} M. Mudrich, S. Kraft, K. Singer, R. Grimm, A. Mosk,
and M. Weidemuller, Phys. Rev. Lett. {\bf 88}, 253001 (2002).

\bibitem{inouye} S.\ Inouye, M.R. Andrews, J. Stenger, H.J. 
Miesner, D.M. Stamper-Kurn and W. Ketterle, Nature \textbf{392}, 151
(1998); 
J.L.\ Roberts, N. R. Claussen, James P. Burke, Jr., Chris H. Greene, E. A. Cornell, and C. E. Wieman, Phys.\ Rev.\ Lett.\ {\bf 81}, 5109 (1998);  
J.\ Stenger, S. Inouye, 
M. R. Andrews, H.-J. Miesner, D. M. Stamper-Kurn, and 
W. Ketterle, Phys.\ Rev.\ Lett.\ {\bf 82}, 2422 (1999); 
S.L.\ Cornish, N. R. Claussen, J. L. Roberts, E. A. Cornell, and C. E. Wieman, 
Phys.\ Rev.\  Lett.\ {\bf 85}, 1795 (2000);

\bibitem{Maxi} G. Izus, M. San Miguel, and M. Santagiustina,
Phys. Rev. E {\bf 64}, 056231 (2001).

\bibitem{stack} P. Kopperschmidt and L. C. Kimerling,
Phys. Rev. B {\bf 63}, 045101 (2001).

\bibitem{PCF} G. Kakarantzas, A. Ortigosa-Blanch, T.A. Birks,
P.S. Russell, L. Farr, F. Couny, B.J. Mangan, Opt. Let. {\bf 28}, 158 (2003).

\bibitem{fetter} A. L. Fetter, J. Low Temp. Phys. {\bf 129}, 263 (2002).

\bibitem{GPE1d} V.M. P\'{e}rez-Garc\'{\i}a, H. Michinel and H. Herrero,
Phys. Rev. A {\bf 57}, 3837 (1998);
 Yu.S. Kivshar, T.J. Alexander and S.K. Turitsyn, Phys. Lett. A {\bf 278}, 225
 (2001); Y.B. Band, I. Towers, and B.A. Malomed, Phys. Rev. A {\bf 67},
023602 (2003).

\bibitem{vpg} J.J. Garc\'{\i}a-Ripoll and V.M. P\'{e}rez-Garc\'{\i}a,
Phys. Rev. A {\bf 63}, 041603 (2001).

\bibitem{thsnbec} D.L. Feder, M.S. Pindzola, L.A. Collins,
B.I. Schneider, and C.W. Clark, Phys. Rev. A
{\bf 62}, 053606 (2000);
B.P. Anderson, P.C. Haljan, C.A. Regal, D.L. Feder, L.A. Collins,
C.W. Clark, and E.A. Cornell, Phys. Rev. Lett.
{\bf 86}, 2926 (2001);
J. Brand and W.P. Reinhardt, Phys. Rev. A {\bf 65}, 043612 (2002).

\bibitem{perez} J.J. Garc\'{\i}a-Ripoll, G. Molina-Terriza,
V.M. P\'{e}rez-Garc\'{\i}a and L. Torner,
Phys. Rev. Lett.  {\bf 87}, 140403 (2001).


\bibitem{JILA}
V. Schweikhard, I. Coddington, P. Engels, V. P. Mogendorff, and
E. A. Cornell, Phys. Rev. Lett. {\bf 92}, 040404
(2004).
\end{references}
\end{document}